\documentclass[doublecol]{epl2}
\usepackage{amsmath}
\usepackage{ulem}

\shorttitle{Is the Tsallis entropy stable?}
\institute{
  \inst{1} Center for Nonlinear Phenomena and Complex Systems CP 231 
Universit\'{e} Libre de Bruxelles, 1050 Brussels, Belgium \\
  \inst{2}Microgravity Research Center, Chimie Physique E.P. CP 165/62,
Universit\'{e} Libre de Bruxelles, Av.F.D.Roosevelt 50, 1050 Brussels,
Belgium.
}

\begin{document}

\title{Is the Tsallis entropy stable?}
\author{James F. Lutsko\inst{1} \thanks{E-mail: \email{jlutsko@ulb.ac.be}} and  Jean Pierre Boon\inst{1} and Patrick Grosfils\inst{2,1}}

\shortauthor{J.F. Lutsko, J.P. Boon and P. Grosfils}


\abstract{The question of whether the Tsallis entropy is Lesche-stable is revisited. It 
is argued that when physical averages are computed with the escort probabilities, the correct application of the concept of Lesche-stability requires use of the escort probabilities. As a consequence, as shown here, the Tsallis entropy is unstable but the thermodynamic averages are stable. We further show that Lesche stability as well as thermodynamic stability can be obtained if the homogeneous entropy is used as the basis of the formulation of non-extensive thermodynamics. In this approach, the escort distribution arises naturally as a secondary structure.}

\maketitle

\section{Introduction}

The concept of non-extensive thermodynamics was introduced by Tsallis about
20 years ago\cite{Tsallis} and has generated a large literature. The
original idea was that for systems out of equilibrium where the Boltzmann
distribution no longer holds, the Boltzmann entropy could be replaced by a
more general function while maintaining the formalism of thermodynamics. In
particular, maximization of the entropy under the usual constraints
(normalized probabilities, fixed internal energy) yields the so-called
q-Gaussian distribution that generalizes the usual Boltzmann distribution of
classical statistical mechanics. While this seems straightforward now, it
did in fact require considerable effort to arrive at the now-accepted form
of the theory. A particular issue that was historically important and that remains problematic is the notion of thermodynamic
stability since the Tsallis entropy gives negative specific heats in certain circumstances\cite{Ramshaw, TsallisPLA, Plastino, AbeNeg}. Nevertheless,
this was one of the issues that motivated the advocation of nontrivial averaging
procedures in non-extensive thermodynamics\cite{TsallisPLA}.

A new controversy has arisen based on a recent paper by Abe
where it is shown that averages computed within the non-extensive formalism
are unstable in the sense that a small change in the distribution function
can lead to a large change in the computed average\cite{AbeEPL}. This
surprising result should be understood in a broader context wherein it was
originally asked whether the Tsallis entropy is stable with respect to
changes in the distribution. This was shown to be true by Abe\cite%
{AbeStability} so the result that averages of observables are unstable while
the entropy is stable appears quite surprising. In this paper we question
whether either fact has actually been proven. In short, our argument is that Lesche-stability
is motivated by making correspondence with an experimental procedure and that
this means it should be understood in terms of the probabilities that govern 
the observation of a given microstate. In the usual formulation of non-extensive thermodynamics,
 those are the escort probabilities. When understood in this way, it is easy to show that the Tsallis
entropy is not Lesche-stable. 

This would appear to create an uncomfortable situation in which the Tsallis formulation
is not Lesche-stable and in which thermodynamic stability is also problematic. We contend that
this can be resolved by a shift of viewpoint in which the physical probabilities are taken
as being fundamental. While the Tsallis entropy cannot be satisfactorily formulated in this way\cite{AbeConcave}, 
a closely related functional, the homogeneous entropy,  appears as a natural alternative. We show
that the homogeneous entropy is in fact Lesche stable, gives positive-definite specific heats, yields 
the usual q-Gaussian distributions when maximized and gives rise to a consistent thermodynamics.

\section{Non-extensive thermodynamics}

\subsection{Tsallis formalism}

The usual non-extensive formalism can be illustrated as follows. Consider a
system composed of some number, $n$, of microstates and let $p_{i}$ be the
probability associated with the $i$-th microstate. The Tsallis entropy is
computed as 
\begin{equation}
S_{q}=\frac{1-\sum_{i=1}^{n}p_{i}^{q}}{q-1}
\end{equation}%
where $q$ is the index characterizing the entropy functional. The limit $%
q\rightarrow 1$ gives the usual Boltzmann entropy. One subtlety in the
theory of non-extensive thermodynamics is that the average of an observable, $%
\mathcal{O}$, that takes the value $O_{i}$ in the i-th state is evaluated
using the so-called escort distribution \cite{BeckSchlogl} giving%
\begin{equation}
\left\langle \mathcal{O}\right\rangle =\sum_{i=1}^{n}\frac{p_{i}^{q}}{%
\sum_{j=1}^{N}p_{j}^{q}}O_{i}  \label{avp}
\end{equation}%
Let the energy observable be $\mathcal{U}$ and let it take on the value $%
\epsilon _{i}$ in the $i$-th state. Then maximization of the entropy under
the constraints of fixed average energy, $\left\langle \mathcal{U}%
\right\rangle =U$, and of normalization, $1=\sum_{i=1}^{n}p_{i}$, gives the $%
q-$Gaussian distribution,%
\begin{equation}
p_{j}=\frac{\left( 1-\left( 1-q\right) Z_{q}^{q-1} \beta\left( \varepsilon
_{j}-U\right) \right) _{+}^{\frac{1}{q-1}}}{\;Z_{q}},\;
\end{equation}%
with%
\begin{equation}
\;Z_{q}=\sum_{j=1}^{N}\left( 1-\left( 1-q\right) Z_{q}^{q-1} \beta \left(
\varepsilon _{j}-U\right) \right) _{+}^{\frac{1}{q-1}}.
\end{equation}%
(Here, the notation $\left( x\right) _{+}$ means $x$ if $x>0$ and zero
otherwise.) It is straightforward to show that if the energy levels are
functions of some parameter, $\varepsilon _{i}=\varepsilon _{i}\left(
\lambda \right) $, then 
\begin{equation}
\frac{dU}{d\lambda }=\beta ^{-1}\frac{dS}{d\lambda }+\frac{dW}{d\lambda }
\end{equation}%
where the work is defined as $dW=\sum_{i=1}^{n}p_{i}d\varepsilon _{i}$. This
is recognized as the first law of thermodynamics.

The peculiar nature of the average, based as it is on the escort
distribution, naturally suggests a reformulation of the theory. If one
defines the new quantity, 
\begin{equation}
P_{i}=\frac{p_{i}^{q}}{\sum_{j=1}^{n}p_{j}^{q}}  \label{pp1}
\end{equation}%
which is invertible,%
\begin{equation}
p_{i}=\frac{P_{i}^{1/q}}{\sum_{j=1}^{n}P_{j}^{1/q}}  \label{pp2}
\end{equation}%
then the Tsallis entropy becomes%
\begin{equation}
\widetilde{S}_{q}=\frac{1-\left( \sum_{i=1}^{n}P_{i}^{1/q}\right) ^{-q}}{q-1}
\label{e2}
\end{equation}%
and averages are computed as normal,%
\begin{equation}
\left\langle \mathcal{O}\right\rangle =\sum_{i=1}^{n}P_{i}O_{i}.  \label{avP}
\end{equation}%
Extremizing this entropy subject to the usual constraints again gives a
q-Gaussian distribution, but with exponent $q/q-1$ rather than $q$. In the
following, we refer to the original, more familiar formulation of
non-extensive thermodynamics in terms of the Tsallis entropy as the
''little-p''\ picture and the reformulation given here as the ''big-P''\
picture. All of this has long been known in the literature of non-extensive
thermodynamics\cite{TsallisThermo}. The little-p formulation is generally
favored because of one important difference: the Tsallis entropy is a
concave function of the probabilities whereas the same is not true of the
big-P entropy. (Concavity is assumed to be required of a generalized entropy
even though the connection between concavity and thermodynamic stability is
complicated by the non-additivity of the entropy\cite{Tatsuaki}.)

On the other hand, there is another important difference between the two
pictures from a more physical point of view: the escort probabilities have
the interpretation of being a measure of the likelihood of finding the
system in a given microstate. This is obvious if one considers an ensemble
of systems in which case the fact that the averages are computed via Eq. (%
\ref{avP}) implies that the fraction of systems in microstate $i$ must be $%
P_{i}$. This is to say that the usual ensemble interpretation of statistical
averages implies a frequentist interpretation of the escort probabilities.
Conversely no such ontology can be imposed on the small-p probabilities.
When using the small-p formulation to express the average $\left\langle 
\mathcal{O}\right\rangle =\sum_{i=1}^{n}\frac{p_{i}^{q}}{%
\sum_{j=1}^{N}p_{j}^{q}}\,O_{i}$, it is clear that $p_i$ does not appear to
measure the frequency of anything. Furthermore, experiments, which measure
averages, are always going to be determining the escort (big-P) distribution
and not the small-p distribution. The fact that both are q-Gaussians means
that this distinction is not of practical importance, but the distinction is
real. The question then arises, is this distinction ever of practical
importance?\ We next show that it plays a critical role in the discussion of
stability of the entropy functional.

\section{Stability}

The discussion of stability is based on ideas first introduced by Lesche
while studying the Renyi entropy\cite{Lesche}. Lesche-stability is defined
in terms of countable families of probability distributions, $W_{n}=\left(
w_{1}^{\left( n\right) },...,w_{n}^{\left( n\right) }\right) $. An entropy
function, $S^{\left( n\right) }\left( W_{n}\right) $ is said to be
Lesche-stable if given two probability distributions, $W_{n}$ and $%
W_{n}^{\prime }=\left( w_{1}^{\left( n\right) \prime },...,w_{n}^{\left(
n\right) \prime }\right) $, for all $\epsilon >0$, there exists $\delta >0$
such that 
\begin{equation}
\delta >\left| W_{n}-W_{n}^{\prime }\right| \equiv \sum_{i}\left|
w_{i}^{\left( n\right) }-w_{i}^{\left( n\right) \prime }\right|  \label{norm}
\end{equation}%
implies that 
\begin{equation}
\epsilon >\left| \frac{S^{\left( n\right) }\left( W_{n}\right) -S^{\left(
n\right) }\left( W_{n}^{\prime }\right) }{S_{\max }^{\left( n\right) }}%
\right|  \label{epsilon}
\end{equation}%
where $S_{\max }^{\left( n\right) }$ is the maximum value possible for the
entropy functional. The quantity $\left| W_{n}-W_{n}^{\prime }\right|$ in
Eq.(\ref{norm}) is the $L_{1}$ norm, and the implication (\ref{norm}) $%
\rightarrow$ (\ref{epsilon}) is required to hold with fixed $\epsilon
,\delta $ in the limit of $n\rightarrow \infty $. This definition is
technical, but the idea is simple: when the the probabilities change, the
change in the entropy should be small if the change in the probabilities is
small. By formulating the definition in terms of a limit, it covers both the
case of probability assignments with finite support (i.e. only some fixed
number of elements are nonzero) as well as assignments with infinite
support. Lesche proved that the Boltzmann entropy is stable, but gave a
simple example showing that the Renyi entropy is not stable\cite{Lesche}.
Abe has given a proof that the Tsallis entropy in the small-p picture is
stable\cite{AbeStability}.

One question that arises immediately is why use the $L_{1}$ norm in
measuring the distance between two distributions? In his original paper\cite%
{Lesche}, Lesche stated that this measure can be related to the difficulty
in experimentally distinguishing two probability assignments. The full
argument was given recently by Abe, Lesche and Mund\cite{LescheAbe} and goes
as follows. To tell whether $W$ or $W^{\prime }$ is the correct description
of a given ensemble of systems, one performs some number, $Z$, of
experiments which, in each instance, determine which microstate the system
is in. Of course, microstates are usually not directly observable, but for
example, one might measure the energy and from this deduce the microstate
(or a set of compatible microstates). The result is a sequence of observed
microstates, $S\equiv (s_{1},...,s_{Z})$. If the correct distribution is $W$%
,the probability to observe this sequence is the product of the
probabilities of observing each microstate, $%
P(s;W)=w_{s_{1}}w_{s_{2}}...w_{s_{Z}}$. Abe et al. define the test of
whether or not the distribution is $W$ by requiring that, prior to the
experiment, the experimenter specify a collection of possible results, $%
C=\left\{ S^{(j)}\right\} $. The probability that the observed $S\in C$ is $%
\alpha =\sum_{j}P\left(S_{j};W\right) $, while the probability $\beta
=1-\sum_{j}P\left( S_{j};W^{\prime }\right) $ is the probability that, if
the true distribution is $W^{\prime }$, the observation will not be in the
set $C$. In order to be an effective test, the set $C$ must be constructed
so that both $\alpha $ and $\beta $ are large:\ i.e., so that if \ $W$ is
correct, the observation is likely to be in $C$ and if $W^{\prime }$ is
correct, the observation is likely to not be in $C$. They show that for
given $Z$, no such test is possible if the $L_{1}$ difference between $W$
and $W^{\prime }$ is too small.

The key observation here is that the relevance of the $L_{1}$ metric is
based on an estimation of probabilities of observing certain outcomes, that
is on the basis of a frequentist measure. Therefore, within the framework of
non-extensive thermodynamics, these probabilities necessarily correspond to
the escort probabilities. Thus, to test Lesche-stability in this formalism,
the $L_{1}$ metric must be applied to the escort probabilities.
Demonstrations based on separation of the small-p distributions measured by
the $L_{1}$ metric do not correspond to the concept of Lesche-stability. An
alternative statement, entirely within the little-p picture, is that one
must check stability of the Tsallis entropy using the metric%
\begin{equation}
\left| p-p^{\prime }\right| =\sum_{i=1}^{n}\left| \frac{p_{i}^{q}}{%
\sum_{j=1}^{N}p_{j}^{q}}-\frac{p_{i}^{\prime q}}{\sum_{j=1}^{N}p_{j}^{\prime
q}}\right| .
\end{equation}

\subsection{The Tsallis entropy is unstable}

So, is the Tsallis entropy Lesche-stable? The answer is that it is not.
Consider the case $q<1$ and let $P$ be the uniform distribution, $P_{i}=%
\frac{1}{n}$ for all $i$, and $P^{\prime }$ be given by $P_{1}^{\prime }=%
\frac{1}{n}+\frac{\delta }{2}$ and $P_{i}^{\prime }=\frac{1}{n}-\frac{\delta 
}{2\left( n-1\right) }$ for $n\geq i>1$. Then, it is obvious that 
\begin{equation}
\left\vert P-P^{\prime }\right\vert _{L_{1}}=\delta
\end{equation}%
and%
\begin{equation}
\widetilde{S}_{q}\left( P\right) =\frac{1-n^{1-q}}{q-1}\,.
\end{equation}%
This happens to be equal to $\widetilde{S}_{q,\max }$. The perturbed
distribution gives%
\begin{eqnarray}
\widetilde{S}_{q}(P') &=&\frac{1-\left( \left( \frac{1}{n}+\frac{\delta }{2}%
\right) ^{1/q}+\left( n-1\right) \left( \frac{1}{n}-\frac{\delta }{2\left(
n-1\right) }\right) ^{1/q}\right) ^{-q}}{q-1} \\
&=&\frac{1-\left( \frac{\delta }{2}\right) ^{-1}\left( 1+{\mathcal{O}}\left( 
\frac{\delta ^{-1}}{n^{1-q}}\right) ^{\frac{1}{q}}\right) }{q-1}  \notag
\end{eqnarray}%
Hence, 
\begin{equation}
\lim_{n\rightarrow \infty }\left\vert \frac{\widetilde{S}_{q}\left( P\right)
-\widetilde{S}_{q}\left( P^{\prime }\right) }{\widetilde{S}_{q,\max }}%
\right\vert =\lim_{n\rightarrow \infty }\left\vert \frac{n^{1-q}-\left( 
\frac{\delta }{2}\right) ^{-1}}{n^{1-q}}\right\vert =1
\end{equation}%
So, no matter how close the distributions (i.e. no matter how small $\delta $%
), the difference between the entropies is finite. There is no difference if
the calculation is translated into the small-p picture.

\section{The homogeneous entropy}

To summarize, our arguments show that the Tsallis entropy with linear
averages is Lesche-stable (as proven by Abe) but with the escort
distribution it is Lesche-unstable. On the other hand, it is known that both
formalisms give negative specific heats giving rise to questions of
thermodynamic stability\cite{Ramshaw,Plastino, Tatsuaki}. In fact, this was one
issue that led to the search for an alternative to linear averages\cite{TsallisPLA}. So, is there a non-extensive formalism that is both
Lesche-stable and that gives positive specific heats? Given the intimate connection
between the concept of Lesche-stability and the physical probabilities, one might
wonder if it would make more sense to use the big-P entropy, Eq.(\ref{e2}) together with
the linear averages as a starting point. As it happens, this is unsatisfactory because of the
fact that the big-P entropy functional is not concave\cite{AbeConcave}. There is, however, a closely
related function known in the information theory literature as the homogeneous entropy\cite{Arimoto} 
given by  
\begin{equation}
S_{q}^{H}\left( P\right) =\frac{\left( \sum_{i=1}^{n}P_{i}^{1/q}\right)
^{q}-1}{q-1}
\end{equation}%
which is concave for all positive $q$, is maximized by the uniform distribution and is extensible\cite{Boekee}. Our proposal
is that this be used in conjunction with the linear averaging procedure as a basis for the formulation
of non-extensive thermodynamics. When maximized under constraint of normalization and fixed internal energy,
the result is a q-Gaussian,%
\begin{equation}
P_{j}=Z_{q}^{-1}\left( 1-\left( 1-q\right) Z_{q}^{1-q}\beta \left(
\varepsilon _{j}-U\right) \right) ^{\frac{q}{1-q}},
\end{equation}%
with%
\begin{eqnarray}
Z_{q} &=&\left( \sum_{i=1}^{n}P_{i}^{1/q}\right) ^{-\frac{q}{1-q}%
} \\
&=&\sum_{i}\left( 1-\left( 1-q\right) \frac{\beta \left( \varepsilon
_{j}-U\right) }{Z_{q}^{q-1}}\right) ^{\frac{q}{1-q}} \notag \\
&=&\sum_{i=1}^{n}\left( 1-\left( 1-q\right) \frac{\beta \left( \varepsilon
_{j}-U\right) }{Z_{q}^{q-1}}\right) ^{\frac{1}{1-q}}  \notag
\end{eqnarray}%
The equality of the second and third lines follows from the constraint on the average energy.
From this expression, it is straightforward to show that 
\begin{equation}
\frac{\partial S_{q}^{H}}{\partial U}=\beta 
\end{equation}%
so that the Lagrange multiplier $\beta $ corresponds to the inverse of the
thermodynamic temperature. Furthermore, the specific heat is%
\begin{equation}
C_{V}=\frac{\partial U}{\partial T}=qZ_{q}^{-\frac{\left( q-1\right) ^{2}}{q}%
}\sum_{i}\left( \beta \left( \varepsilon _{j}-U\right) \right) ^{2}P_{i}^{%
\frac{2q-1}{q}}
\end{equation}%
which is positive definite.  It is also easy to show that if the energies $%
\varepsilon _{i}$ are a function of some external parameter, $\lambda $, then%
\begin{equation}
\frac{\partial U}{\partial \lambda }=\beta ^{-1}\frac{\partial S_{q}^{H}}{%
\partial \lambda }+\sum_{i=1}^{n}p_{j}\frac{\partial \varepsilon _{j}}{%
\partial \lambda }
\end{equation}%
which confirms the second law of thermodynamics for this model.

Lesche-stability is also easy to show. Let $\delta _{i}=P_{i}^{\prime }-P_{i}
$, $\sum \left\vert P_{i}^{\prime }-P_{i}\right\vert =\delta <1$ and assume
without loss of generality that $S\left( P^{\prime }\right) >S\left(
P\right) $ where $S\left( P\right) \equiv \left( q-1\right)\left({S_{q}^{H}\left( P\right) +1}\right)$. Then, for $q<1$, one has that 
\begin{eqnarray}
S\left( P^{\prime }\right)  &=&\left( \sum_{i=1}^{n}\left( P_{i}+\delta
_{i}\right) ^{1/q}\right) ^{q} \\ &=&\left( \sum_{i=1}^{n}\left\vert P_{i}+\delta
_{i}\right\vert ^{1/q}\right) ^{q} \notag \\
&\leq &\left( \sum_{i=1}^{n}P_{i}^{1/q}\right) ^{q}+\left(
\sum_{i=1}^{n}\left\vert \delta _{i}\right\vert ^{1/q}\right) ^{q}  \notag
\end{eqnarray}%
The last line follows from the Minkowski inequality (a generalization of the triangle inequality)\cite{Minkowski}. For $q<1$ and $%
\left\vert \delta _{i}\right\vert \leq \delta <1$, one has $\left\vert
\delta _{i}\right\vert ^{1/q}<\left\vert \delta _{i}\right\vert $ so%
\begin{equation}
S\left( P^{\prime }\right) \leq \left( \sum_{i=1}^{n}P_{i}^{1/q}\right)
^{q}+\left( \sum_{i=1}^{n}\left\vert \delta _{i}\right\vert \right)
^{q}=S\left( P\right) +\delta ^{q}.
\end{equation}%
Lesche-stability follows immediately. For $q>1$ the proof is slightly more
complicated. Note that if $x\geq y>0$ and $q>1,$ then $q\left( x-y\right)
x^{q-1}\geq x^{q}-y^{q}$, which follows from the fact that $q\left(
x-y\right) x^{q-1}-x^{q}+y^{q}$ is monotonically decreasing as a function of 
$y$. Making all the same assumptions as above, this implies that 
\begin{eqnarray}
\left( \sum_{i=1}^{n}P_{i}^{\prime 1/q}\right) ^{q}-\left(
\sum_{i=1}^{n}P_{i}^{1/q}\right) ^{q} \\
 \leq  q\left( \sum_{i=1}^{n}\left(
P_{i}^{\prime 1/q}-P_{i}^{1/q}\right) \right) \left(
\sum_{i=1}^{n}P_{i}^{\prime 1/q}\right) ^{q-1}. \notag
\end{eqnarray}%
Next, note that for $q>1$ and $x>y>0$, $x^{1/q}-y^{1/q}\leq \left(
x-y\right) ^{1/q}$ as follows from the fact that $\left( x-y\right)
^{1/q}-x^{1/q}+y^{1/q}$ is convex as a function of $y$. Thus%
\begin{eqnarray}
\left( \sum_{i=1}^{n}P_{i}^{\prime 1/q}\right) ^{q}-\left(
\sum_{i=1}^{n}P_{i}^{1/q}\right) ^{q} \\
\leq q\left( \sum_{i=1}^{n}\left\vert
\delta _{i}\right\vert ^{1/q}\right) \left( \sum_{i=1}^{n}P_{i}^{\prime
1/q}\right) ^{q-1}. \notag
\end{eqnarray}%
The two sums on the right can be bounded by maximizing  $\sum_{i=1}^{n}x_{i}^{1/q}$ subject to $%
\sum_{i=1}^{n}x_{i}=\gamma $ using a Lagrange multiplier. The result is that  $x_{i}=\gamma/n$ and the sum is $\gamma^{1/q}n^{(q-1)/q}$ giving 
\begin{eqnarray}
\left( \sum_{i=1}^{n}P_{i}^{\prime 1/q}\right) ^{q}-\left(
\sum_{i=1}^{n}P_{i}^{1/q}\right) ^{q} \\
\leq q\delta ^{1/q}\left(
n^{(q-1)/q}\right) ^{q}=q\delta ^{1/q}n^{q-1} \notag
\end{eqnarray}%
which, after normalization, implies Lesche-stability.

Finally, we can make contact with the Tsallis entropy as follows. The
normalization condition can be enforced by eliminating one of the degrees of
freedom, i.e. by using%
\begin{equation}
P_{n}=1-\sum_{i=1}^{n-1}P_{i}
\end{equation}%
However, this is awkward as it treats one degree of freedom differently
from the others. A more symmetrical way to impose it would be to introduce
auxiliary quantities, $u_{i}$, and to write%
\begin{equation}
P_{i}=\frac{u_{i}}{\sum_{j=1}^{n}u_{j}}
\end{equation}%
This is completely general. Note that it is degenerate as $\left\{
u_{i}\right\} $ and $\left\{ \lambda u_{i}\right\} $ give the same $\left\{
p_{i}\right\} $. Since $p_{i}>0$, all of the $u_{i}$ must be either positive
or negative. Again, without loss of generality, we can take them all to be
positive. In terms of these, the entropy becomes%
\begin{equation}
S_{q}^{H}\left( p\right) =\frac{\left( \sum_{i=1}^{n}u_{i}^{1/q}\right)
^{q}\left( \sum_{j=1}^{n}u_{j}\right) ^{-1}-1}{q-1}
\end{equation}%
We can simplify this by using the freedom mentioned above to impose the
single constraint,%
\begin{equation}
1=\sum_{i=1}^{n}u_{i}^{1/q}
\end{equation}%
which is equivalent to choosing a particular scaling of the $u$'s. Note that
since all the $u_{i}>0$, this implies that they are all also less than one.
Then, it appears to be more convenient to introduce $v_{i}=$ $u_{i}^{1/q}\in %
\left[ 0,1\right] $ giving%
\begin{align}
P_{i}& =\frac{v_{i}^{q}}{\sum_{j=1}^{n}v_{j}^{q}} \\
S_{q}^{H}\left( v\right) & =\frac{\left( \sum_{j=1}^{n}v_{j}^{q}\right)
^{-1}-1}{q-1}=\frac{1-\sum_{j=1}^{n}v_{j}^{q}}{\left( q-1\right)
\sum_{j=1}^{n}v_{j}^{q}}  \notag \\
1 & =\sum_{i=1}^{n}v_{i}  \notag \\
\left\langle \mathcal{O}\right\rangle & =\sum_{i=1}^{n}\frac{v_{i}^{q}}{%
\sum_{j=1}^{n}v_{j}^{q}}O_{i}  \notag
\end{align}%
The "escort probabilities" therefore arise naturally as a way of encoding
the normalization constraint. However, it is clear in this interpretation
that the $\left\{ v_{i}\right\} $ are not physical probabilities but just
quantities that happen to be positive and to sum to one. The form of the homogeneous entropy written in terms of
the $\left\{ v_{i}\right\} $ is known in the literature and is called the
"normalized\ Tsallis entropy"\cite{AbeNormalized}. However, note that the normalized
Tsallis entropy is only concave for $q \le 1$ so the homogeneous
entropy is a more general starting point. 

\section{Concluding Comments}

Our conclusion is that when the concept of Lesche stability is properly
applied within the usual formalism of non-extensive thermodynamics, the
Tsallis entropy is just as unstable as the Renyi entropy originally
considered by Lesche. Whether or not this conclusion, based on particular admittedly artificial
examples, is physically relevant and would have practical implications is a
question of on-going debate \cite{TsallisThurner} just as was the question
of the stability of the Renyi entropy \cite{LeschePRE}. However, arguments
such as those given in \cite{TsallisThurner} should be reconsidered in light
of the correct application of the $L_{1}$ measure.

A further consequence which follows from the big-P formulation concerns the
stability of the averages of observables. In a recent demonstration they
were shown to be unstable in the non-extensive formalism \cite{AbeEPL}. This
result was obtained using the small-p picture where the interpretation of
the probabilities is problematic, while it was shown that in the classical
formulation of statistical mechanics with linear averages, stability was
guaranteed. This is precisely in accordance with our arguments: since the $%
L_1$ measure should be applied to the escort probabilities (i.e. the big
P's), it follows that the averages are stable.

In summary, the use of the $L_1$ norm in framing the concept of
Lesche-stability is justified by considering an experimental test designed
to distinguish different hypothesized probability distributions. As such, it
is a property not only of the form of the entropy but also of the means used
to relate the ``probabilities'' occuring in the entropy to experiment. Our
conclusion is that the result of Abe\cite{AbeStability} should be
interpreted as demonstrating that the combination of Tsallis entropy and
linear averages is stable while the combination of Tsallis entropy with the
escort distribution averages is unstable (as shown by our example above).
This being the case, the further observations concerning the stability of
the averages\cite{AbeEPL} only reinforces these conclusions.

On the other hand, we have shown that the homogeneous entropy with the usual
linear averaging procedure provides a satisfactory starting point for the development
of non-extensive thermodynamics. We note that this is not the first time the homogeneous entropy has occured in 
the context of non-extensive thermodynamics. For example, Lavenda and Dunning-Davies have used it as an example
illustrating that certain features of the Tsallis entropy are not unique\cite{Lavenda}. In the form of the normalized Tsallis distribution, it 
appears to have first been discussed by Rajagopal and Abe\cite{AbeNormalized} where it was noted that it is only concave for $q \in [0,1]$.
Abe subsequently concluded that it is not Lesche-stable, but by our interpretation, this argument shows instability when the linear averaging procedure
is used\cite{AbeStability}. We also note that Lenzi et al  had already demonstrated that the normalized Tsallis entropy gives positive
specific heats\cite{Lenzi}. Our main contribution has been to note that the homogeneous entropy is more general than the normalized Tsallis entropy (because it is concave for all values of $q$) and to show it is Lesche-stable. Together with other properties, such as the positivity of the specific heats, we suggest
this makes it a preferred starting point for the development of non-extensive thermodynamics.

Finally, this approach also sheds light on the underlying ontology of the formalism. The concept of the escort distribution
has become an accepted part of the non-extensive formalism but, as we have discussed, it is hard to understand in what sense
the small-p quantities are ``probabilities'' as opposed to a set of quantities that happen to be positive and normalized. Nevertheless, 
various arguments have also been given for using the escort-distribution rather than the linear averaging procedure\cite{AbeGeometry,AbeNecessity}. In our formulation, this ambiguity is clarified: the only \emph{probabilities} are the physical probabilities used to compute averages. The escort distribution
arises as a natural way to simplify the form of the entropy, but the equivalent of the small-p variables are clearly quantities to which
no physical significance attaches. 

\begin{acknowledgments}
We thank Sumiyoshi Abe for useful discussions. 
The work of JFL was supported by the European Space Agency under contract
number ESA AO-2004-070 and that of PG by the projet ARCHIMEDES of the Communaut\'e
Fran\c caise de Belgique (ARC 2004-09). 
\end{acknowledgments}

\bigskip

\bibliographystyle{eplbib}

\end{document}